\documentclass{article}

\usepackage[utf8]{inputenc}
\usepackage[margin = 1in]{geometry}

\usepackage{graphicx} 
\usepackage{makecell}

\usepackage{dsfont}
\usepackage{hyperref}
\usepackage{amsmath}
\def\b{\ensuremath\mathbf}
\usepackage{appendix}
\usepackage{booktabs}
\usepackage{graphicx}
\usepackage{multirow}
\usepackage{siunitx}
\usepackage{gensymb}

\usepackage{amsmath}
\usepackage{bm}
\usepackage{graphicx}
\usepackage{xcolor}
\usepackage{hyperref}
\usepackage{dsfont}
\def\b{\ensuremath\mathbf}

\usepackage{soul}

\usepackage{graphicx}
\usepackage{dcolumn}
\usepackage{bm}

\usepackage[utf8]{inputenc}
\usepackage[T1]{fontenc}
\usepackage{mathptmx}
\usepackage{etoolbox}

\usepackage{xcolor}
\usepackage{ dsfont }

\usepackage{hyperref}
\usepackage{amsfonts}

\def\b{\ensuremath\mathbf}

\title{A  data-driven biophysical  network model\\
reproduces \textit{C. elegans} premotor neural dynamics}

\author{Megan Morrison$^1$ and Lai-Sang Young$^2$\\[.1in]
$^1$ Applied Mathematics, Illinois Institute of Technology, Chicago, IL 60616\\
$^2$ Courant Institute of Mathematical Sciences, New York University, New York, NY 10012
}

\begin{document}

\maketitle

\begin{abstract} 

\normalsize

\textit{C. elegans} locomotion is composed of switches between forward and reversal states 
punctuated by turns.  This locomotory capability is necessary for the nematode to move towards attractive stimuli, escape noxious chemicals, and explore its environment. Although experimentalists have identified a number
of premotor neurons as drivers of forward and reverse motion, how these neurons work together to produce the behaviors observed remains to be understood. Towards a better understanding of \textit{C. elegans} neurodynamics, we present in this paper a minimally parameterized, biophysical dynamical systems model of the premotor network.
Our model consists of a recurrently connected collection of premotor neurons (the core group) driven by over a hundred sensory and interneurons that provide diverse feedforward inputs to the core group. 
It is data-driven in the sense that the choice of neurons in the core group follows experimental guidance, anatomical structures are dictated by the connectome, 
and physiological parameters are deduced from whole-brain imaging and voltage clamps data.
When simulated with realistic input signals, our model produces premotor activity that closely resembles experimental data:  from the seemingly random switching between forward
and reversal behaviors to the synchronization of subnetworks to various higher-order statistics. 
 We posit that different roles are played by gap junctions and synaptic connections in switching dynamics. The model correctly 
predicts behavior such as dwelling versus roaming as a result of the synaptic inputs received,
and we demonstrate that it can be used to study how the activity level of certain
individual neurons impacts behavior.
\end{abstract}

\subsection*{Author Summary}
\large

The nematode \textit{C. elegans} has one of the most complete connectomes and relatively
few behavioral states, making it an ideal organism for which to attempt the formidable task
of connecting biology to function. This paper is about the premotor
network, which controls locomotion. We model premotor activity as a driven dynamical
system: a recurrent network of $\sim$15 neurons known to be intimately connected
to forward and reversal movements, driven by over 100 of their presynaptic neurons. 
Whole-brain imaging data are used to constrain parameters. Comparing model outputs to data, we show that the model accurately reproduces basic features of \textit{C. elegans} movements, such as the seemingly random switches 
between forward crawling and reversals. Because our model of neural activity
is semi-realistic and analyzable, 
it has the potential to reveal mechanisms and predict behavior. As an example, we use it to clarify the role of gap junctions versus synaptic inputs and the dependence on various sensory neurons.

\section*{\label{sec:level1}Introduction}

To survive, animals must be able to generate a wide range of behaviors and to switch between them effectively.  These behaviors, ubiquitous across species, include searching for food, forming social aggregates, and escaping from danger.
The neural mechanisms responsible for producing different behaviors and inducing transitions are not well understood, even in the simplest of organisms.
\textit{C. elegans}, with their relatively simple nervous system, limited behavioral states, 
and amenability to experimentation \cite{sengupta_caenorhabditis_2009, sattelle_invertebrate_2006, lin_imaging_2022}, are an ideal candidate to attempt a theory 
that connects neurobiology to behavior. The building of such a theory
is an overarching goal of the present work.

The \textit{C. elegans} connectome 
consists of 300+ neurons; interneurons act as the main processors of information, they receive input from sensory neurons and send 
commands to motor neurons \cite{sabrin_hourglass_2020, roberts_stochastic_2016, kaplan_sensorimotor_2018}.  To some extent, experimentalists have demystified 
how sensory neurons respond to stimuli and how motor neurons coordinate 
muscle activity \cite{goodman_how_2019, bargmann_chemosensation_2006, zhen_c_2015, wen_proprioceptive_2012, gao_excitatory_2018}, but relatively little is known about
how the intermediary interneurons operate collectively to make behavioral decisions.
A subset of interneurons --- premotor neurons --- are chiefly responsible for determining the most common locomotory behaviors \cite{roberts_stochastic_2016, kaplan_sensorimotor_2018, zhen_c_2015}.
In this work, we model and analyze the activity of the premotor network of \textit{C. elegans}.

Despite the remarkable progress that has been made in documenting
various aspects of \textit{C. elegans} neural circuits, gaps in our knowledge remain.
Individual neurons that promote forward and reversal movements have been identified, but the mechanisms that govern their collective activity remain to be understood.
The connectome \cite{white_structure_1986} provides detailed, quantitative information on connections 
between pairs of neurons in the entire organism, but anatomy alone -- without 
physiology or dynamics -- does not predict activity. Whole-brain imaging 
data \cite{kato_global_2015, atanas_brain-wide_2023} connects neuronal activity to behavior, yet does not explain how activity patterns are generated.

We propose that computational modeling may supply some of the missing information,
and that models with the following properties could be useful:
(i) The model should have a clear 
representation of anatomical structures and physiological measurements; without these components,
model inferences are hard to interpret.
(ii) The model should be biophysical.  
(iii) Dynamics being one of the missing ingredients, it would be desirable to have
a dynamical systems model capable of producing the range of activity and subsequent behaviors
observed. (iv) The model's responses must be realistic, i.e. its activity outputs 
must be similar to data.  (v) Lastly, the model must be
analyzable, as models are built to be interrogated.

We present in this paper a model of the \textit{C. elegan} premotor network aimed at
(i)--(v) above. We focus on a core group of 10-20 neurons that experimentalists have
identified as drivers of forward and reversal movements and model the dynamical interaction
among them by a system of differential equations.  The state variables are the neurons' calcium levels. This core group, which is small enough to be analyzable, receives inputs 
from over 100 presynaptic 
neurons. We do not model the dynamics of the presynaptic neurons.  We instead treat their influence as an external force, i.e., we model premotor activity as a {\it driven 
dynamical system}. Leaving model details to the \nameref{sec:results}, we remark that
we have leaned heavily on connectomic data for network architecture, and on many sets of
whole-brain imaging data for model parameters.

A signature of \textit{C. elegans} locomotion is its seemingly random switching between
forward crawling and reversals. These switches are, in fact, far from random: they enable the 
\textit{C. elegans} to forage for food and linger in a favorable environment. 
When provided with realistic input, our model can reproduce salient characteristics of switching dynamics.  It does so through partial synchronization
of clusters of neurons similar to the activity seen in the data. The dynamics produced
are complex: they are neither completely random (as in the Markov chains models \cite{roberts_stochastic_2016, linderman_hierarchical_2019} nor are they limit cycles (as in earlier dynamical systems models \cite{lanza_recurrent_2021, kunert_spatiotemporal_2017}). 
 The close resemblance of our model outputs to data enhances the plausibility
of our findings.
Model analysis reveals, among other things, the differing roles played by gap junctions and synaptic connections in switching dynamics.
 
Building models with properties (i)--(v) is not without challenges. We will discuss some of the issues, 
model limitations and relations to existing work in the \nameref{sec:discussion} section.

 \section*{Results}\label{sec:results}

We fit a dynamical systems model of premotor interneurons using several sources of data -- voltage clamps, whole-brain calcium imaging, and the connectome. More precisely, we use the known connectome structure to set the locations of gap junction and synaptic connections and then use the whole-brain imaging data in conjunction with voltage clamp data to determine parameters, which consist primarily but not exclusively of the signs and weights of synaptic connections. 
Our model is a recurrent network of premotor neurons driven by a large collection of
their presynaptic neurons; the presynaptic activity is also taken from the imaging data.

Our main results can be summarized as follows: 
\begin{itemize}
\item[(1)]  Our model reproduces stochastic switching between the forward and reversal premotor neuron clusters, resembling data.
\item[(2)] The model reproduces neuronal time series whose higher-order statistics match those observed in the data.  
\item[(3)]Using the deduced synaptic sign estimates, we identify presynaptic neurons that promote specific behaviors.
\item[(4)] We use our model to show how gap junctions and synaptic connections may collaborate to produce emergent stereotyped activity in the premotor network.
\end{itemize}


\subsection*{Model description}

\begin{figure}[th!]
   \centering
\includegraphics[width = 0.8\linewidth]{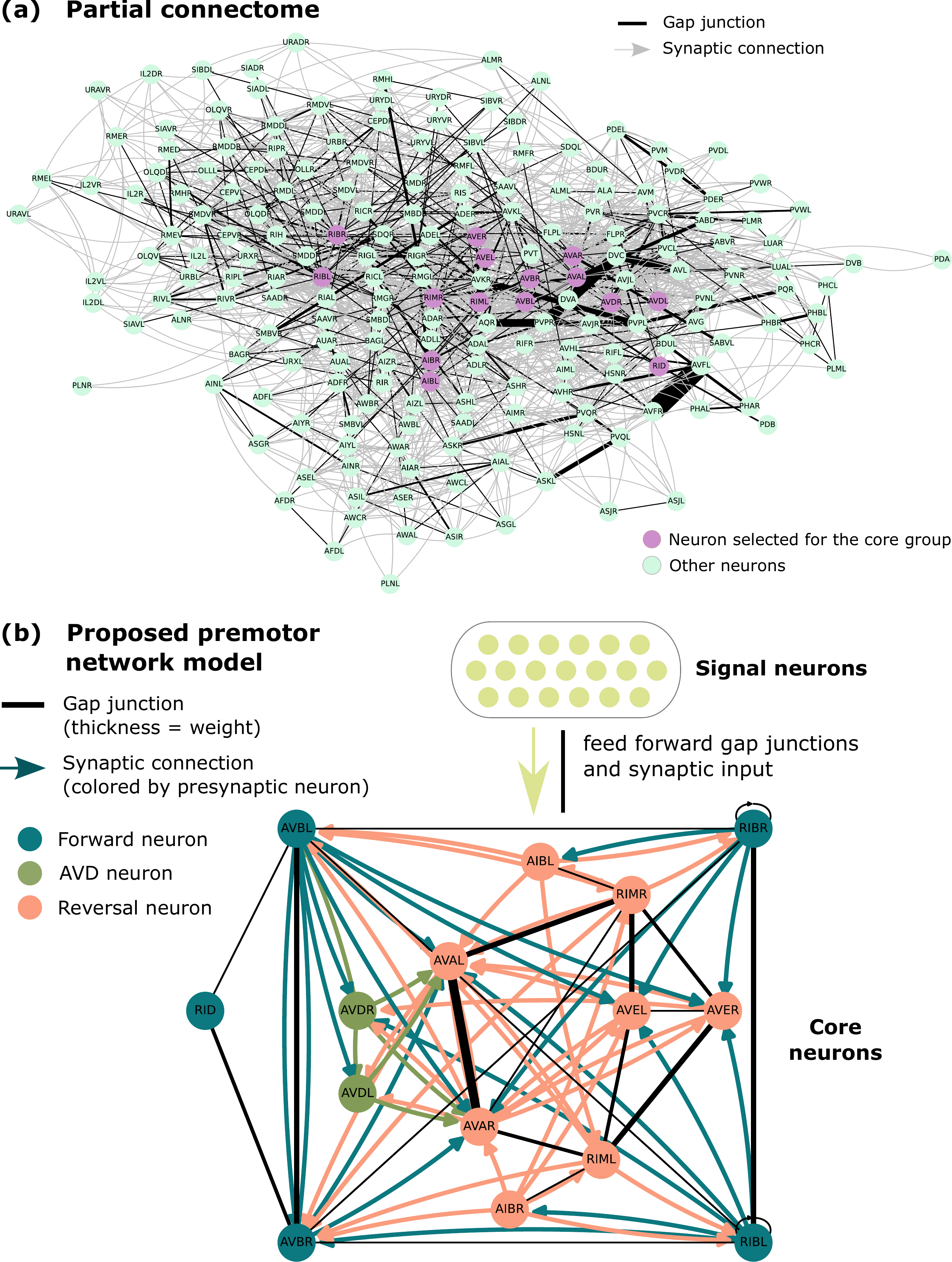}
   \caption{\fontsize{10}{12}\selectfont (a) Partial connectome from \cite{white_structure_1986} containing 198 sensory neurons, interneurons, and motor neurons (most motor neurons that form neuromuscular junctions with muscle cells are excluded).  The 15 neurons that will be selected as core neurons are shown in purple.  Additionally, the partial connectome contains 137 neurons that are directly connected to the core neurons either presynaptically or through gap junctions, and 46 neurons that are connected through an intermediary neuron. (b)  Core neurons selected for the premotor network model and the connections among them.  We categorize most core neurons as either forward or reversal based on the connectome and previous experimental work.  Core neurons receive input from signal neurons that are presynaptic to the core neuron set or connected via gap junctions. Highly connected signal neurons shown in Figure~\ref{fig:syn_weights_avgs}.}
   \label{fig:connectome_premotor}
\end{figure}

To build a dynamical network of premotor neurons, we must first decide which neurons to include in the model.
Experimental and whole-brain imaging studies have
identified that various neurons are associated with forward and reversal locomotion, but
there is no obvious way to isolate a subgraph of the connectome that corresponds
to {\it the premotor network}. 
Premotor neurons receive inputs from many sensory neurons and interact with
many other interneurons that may or may not be directly implicated in locomotion; these neurons, in turn, interact with a larger set of neurons, and so on. If all potentially
relevant neurons were included, 
the resulting network
would be exceedingly large; see Figure~\ref{fig:connectome_premotor}(a). 
Such a model would be impossible to fit due to a deficit of labeled neurons in whole-brain imaging data and difficult to analyze due to a large number of variables and parameters.

Attempting to strike a balance between biological realism and analyzability, we
propose a model with the following basic structure: a core group of neurons whose dynamics we simulate while influenced by a larger group of neurons whose activity is applied as input. 
More precisely, neurons in our network are divided into two groups. The first is a core group of neurons deemed the most
relevant to the study in question. For us, these would consist of interneurons believed to play the most important roles in \textit{C. elegans} locomotion.
Neurons in this group are coupled to one another in a recurrent network; 
we will refer to them as {\it core neurons} for simplicity.  This group should be
relatively modest in size, though as large as need be.
Then there is a second group consisting of 
 what we will refer to as  {\it signal neurons}. Signal neurons provide feedforward 
 input to the core group;  we do not model 
 their dynamics, gleaning their outputs instead from various data sets. 
Core neurons are assumed to only receive input from other core neurons and signal neurons.

Mathematically, we build a model in the form of a {\it driven dynamical system},
 defined by a set of ordinary differential equations  describing the dynamics of
 neurons in the core group, together with a time-dependent forcing representing 
 the inputs from signal neurons. 
Below we identify the neurons to be modeled and give
the equations governing model dynamics.


\paragraph{\large Selection of neurons}
Guided by connectomic, whole-brain imaging, and experimental data, and constrained by data availability, (\cite{white_structure_1986, atanas_brain-wide_2023, kato_global_2015, kaplan_sensorimotor_2018, zhen_c_2015}), 
we converge to a set of 15 neurons that, we posit, play central roles in 
forward-reversal locomotion. These are the neurons we place in the core group in our model. 
Below we recall some basic biological facts about them, and explain why they were chosen.

Conceptually, we cluster the 15 core neurons into three categories: 
(i) {\it Forward neurons} (AVBL, AVBR, RIBL, RIBR, and RID); the AVB neurons are known
to be command premotor neurons that drive forward movement;  they are connected
to the other interneurons in the group via gap junctions \cite{white_structure_1986, zhen_c_2015, roberts_stochastic_2016, kaplan_sensorimotor_2018, rakowski_optimal_2017, riddle_mechanosensory_1997, uzel_set_2022, flavell_behavioral_2020}. 
(ii) {\it Reversal neurons} (AVAL, AVAR, RIML, RIMR, AVEL, AVER, AIBL, and AIBR);
the AVA neurons are known to be command premotor neurons that drive reversal 
movements;  they are connected to the other interneurons in the group via gap junctions \cite{lanza_recurrent_2021, gray_circuit_2005}.
The forward and reversal clusters above have many synaptic connections within and 
between clusters. Finally, (iii) there are two neurons (AVDL and AVDR), which are highly 
interconnected with the two clusters via synaptic connections. A diagram showing
these 15 neurons together with their known connections is shown in Figure~\ref{fig:connectome_premotor}(b).

According to the connectome, 137 neurons provide input to the core group \cite{white_structure_1986}; 112 of the 137 input neurons appear in the whole-brain imaging data used to fit our model \cite{atanas_brain-wide_2023} --- these are the 112 signal neurons defined in our model. 
Divergent from our model, in the real \textit{C. elegans} brain the 15 core neurons have many recurrent connections with the signal neurons --- we have chosen to neglect these recurrent interactions in favor of a more tractable dynamical system.

We remark that though guided by data, there is some arbitrariness in the choices 
above as well as practical constraints.  Most notably, we do not include the PVC and DVA neurons in our core set despite their strong association with locomotion and strong connections with the core set.  This is because the PVC and DVA neurons do not appear in the whole-brain imaging datasets we use to fit our model \cite{atanas_brain-wide_2023}, making it infeasible to fit model parameters for these neurons using the methods we employ.  Another omission of note is that we exclude motor neurons, many of which
provide feedback to our core group via gap junctions \cite{white_structure_1986}.
As these neurons are ``downstream" from our network, we do not wish to model them as signal neurons.

\begin{figure}[th!]
   \centering
   \includegraphics[width = 1.0\linewidth]{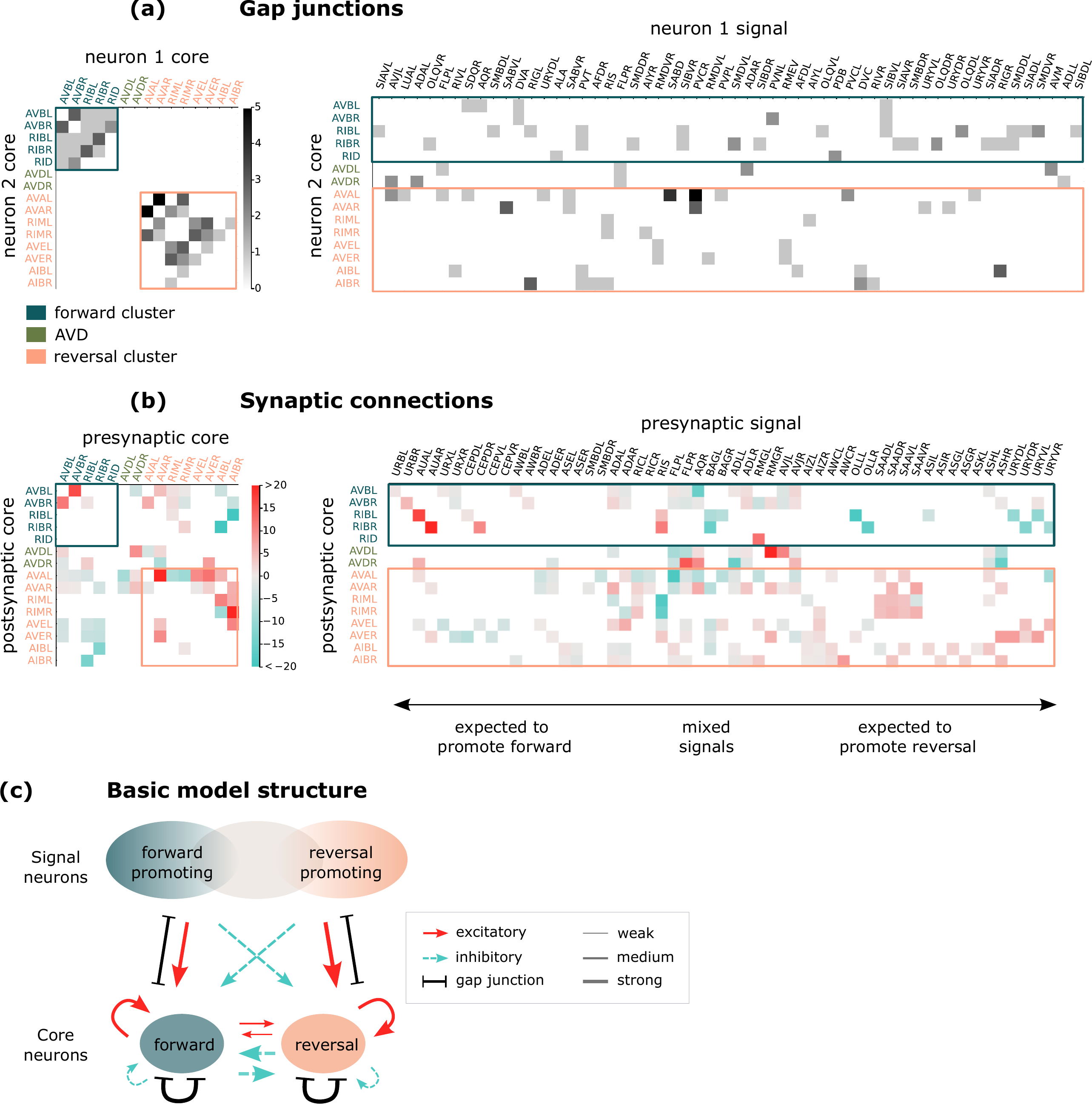}
   \caption{(a)  Relative gap junction weights between core neurons and between signal neurons and core neurons (select signal neurons are shown). Weights are taken directly from \cite{white_structure_1986}. 
   (b)  Synaptic signed weights between core neurons and between signal neurons and core neurons derived from regression
   (select signal neurons are shown).  Signal neurons are loosely sorted as forward-promoting or reversal-promoting.  (c)  Cartoon of principal structure between signal neurons, forward core neurons, and reversal core neurons.}
   \label{fig:syn_weights_avgs}
\end{figure}

\paragraph{\large Model dynamics and parameters}
We begin with model equations for a generic
set of $n$ core neurons and $m$ signal neurons. Conditions having to do with the 
specific core neurons selected are imposed only at the end.

As \textit{C. elegans} neurons have graded responses to injected currents, it is natural 
to measure their activity in terms of calcium levels \cite{goodman_active_1998}. 
We set the state variable for each neuron to be its GCamp z-score; these measurements are acquired from whole-brain imaging data \cite{atanas_brain-wide_2023}.
 Let $x_1, \cdots, x_n$ denote these scores for the
$n$ core neurons and $x_{n+1}, \cdots, x_{n+m}$ the scores for the signal neurons.
We posit that for $i \le n$, the time evolution of $x_i$ is given by
\begin{align} \label{eq:general_model-results}
\tau \frac{dx_i}{dt} &=  f_i(x_i) + \beta \sum_{j=1}^{n+m}\b{W}_{ij}(x_j - x_i) +\sum_{j=1}^{n+m}\b{A}_{ij} \sigma(x_j).
\end{align}
The first term on the right side describes neuron $i$'s intrinsic dynamics and the second term  
represents the combined influence from all core neurons $j$ that form 
gap junctions with neuron $i$. The third represents the influence from
core neurons presynaptic to neuron $i$; $\tau, \beta, \b{W}_{ij}, \b{A}_{ij}$ are
constants, and $\sigma() = \text{ReLU}()$ is the ReLU activation function.

As stated earlier, the dynamics of $x_i$ for $i>n$, i.e., for signal neurons,
will not be modeled. Instead, $\{x_i(t)\}$ is extracted from whole-brain imaging time series.

The intrinsic dynamics of individual neurons are approximated by 
voltage clamp data \cite{goodman_active_1998, nicoletti_biophysical_2019}. 
Leaving details to the \hyperref[method:intrin_d]{Methods}, we assume $f_i(x_i)$ has the form 
$$f_i(x_i) = -2(x_i+0.8)(x_i-0.1)(x_i-1)  + \b{d}_i$$  
where the bias term $ \b{d}_i$ is different for each neuron. In the absence of input 
from other neurons and without a bias term, the system has two stable states,  a low-activity state at $x_i = -0.8$ 
and a high-activity state at $x_i = 1$. A positive bias $\b{d}_i$ makes the neuron's 
high-activity state relatively more stable while a negative $\b{d}_i$ makes the neuron's 
low-activity state more stable. The cubic form of the neurons' intrinsic dynamics 
approximates the typical nonlinear response of \textit{C. elegans} neurons to current injection \cite{goodman_active_1998, nicoletti_biophysical_2019}.

The influence from both gap junctions and synaptic connections is approximated as linear,
but the latter involves a rectifier activation function. The rectifier activation function 
applied to the neural activity of the presynaptic neuron reflects our assumption that neurons 
influence their postsynaptic counterparts synaptically only when depolarized and do 
not affect postsynaptic neurons when hyperpolarized.

\medskip
The quantities that remain to be determined are $\b{d}_i$ (bias in the intrinsic dynamics of
individual neurons), $\b{W}_{ij}$ (relative gap junction weights between neurons), $\b{A}_{ij}$
(synaptic weights), $\beta$ (relative contribution of gap juction input), and $\tau$ (timescale parameter). These quantities depend on the specific neurons in the model, and here
is how they are determined:

The gap junction weights $\b{W}_{ij}$ are derived from connectome data \cite{white_structure_1986} (see \hyperref[method:gap_j]{Methods});
for the neurons in our model, they are shown in Figure~\ref{fig:syn_weights_avgs}(a).

As for the synaptic weights $\b{A}_{ij}$, we set $\b{A}_{ij}=0$ 
when the connectome shows no connection between the relevant neurons;
absent connections are depicted by white boxes in Figure~\ref{fig:syn_weights_avgs}(b). 
When a connection is present, connectome data offer information on the number of
synapses {\it without sign specification} \cite{white_structure_1986}. Interpretation of this data is further complicated
by the fact that many synaptic connections 
within the locomotion subnetwork are known to be ``complex", meaning the connection 
contains both excitatory and inhibitory neurotransmitter-receptor pairs \cite{fenyves_synaptic_2020}. 
In the absence of clear guidance, we have elected to estimate independently 
the values of  $\b{A}_{ij}$, which in our model 
can be positive or negative representing the {\it net} excitatory/inhibitory inputs.

The $\b{A}_{ij}$ not set to $0$ {\it a priori}, the values of $\b{d}_i$, and $\beta$ are estimated by linear regression. 
We first find datasets from \cite{atanas_brain-wide_2023} that contain a sufficient quantity of core neurons, resulting in 22 datasets that we use to fit our model. We then perform a parameter sweep for $\beta$ over many linear regressions.  
The minimum combined error occurs when $\beta = 10$ (See \hyperref[method:param_fit]{Methods}).
After setting $\beta$ to be $10$, we perform separate linear regressions for each of the 22 datasets and average over all of the dataset parameters to obtain a single general set of parameter values $\b{A}_{ij}$ and $\b{d}_i$.
The obtained values for the synaptic weights and signs, $\b{A}_{ij}$, are shown in Figure~\ref{fig:syn_weights_avgs}(b).

The last parameter that must be fit is the timescale parameter $\tau$.  We fit $\tau$ by simulating the activity of the core neurons in the six datasets shown in Figure~\ref{fig:simulation_data_match} for variable $\tau$ values.
We select $\tau$ such that, on average, the simulated core neurons
switch at the same speed as the real neurons (see \hyperref[methods:sim]{Methods}, Figs.~\ref{fig:timescale_increase}, \ref{fig:simulation_data_match}).

Recapitulating, the neurons in our core group contain two subsets --- the "forward cluster" and the "reversal cluster" --- that are known from experiments to be associated with forward and reversal locomotion. (See the paragraph on neuron selection).
Simulated dynamics of these core neurons will be 
studied and compared to data. We note that in our model, dynamical properties of individual 
neurons are emergent; the only information specific to a neuron that we have built into the model
are gap junction weights and the presence or absence of synaptic connections.

 \subsection*{Model validation} 
 
\begin{figure}[th!]
   \centering
   \includegraphics[width = 0.8\linewidth]{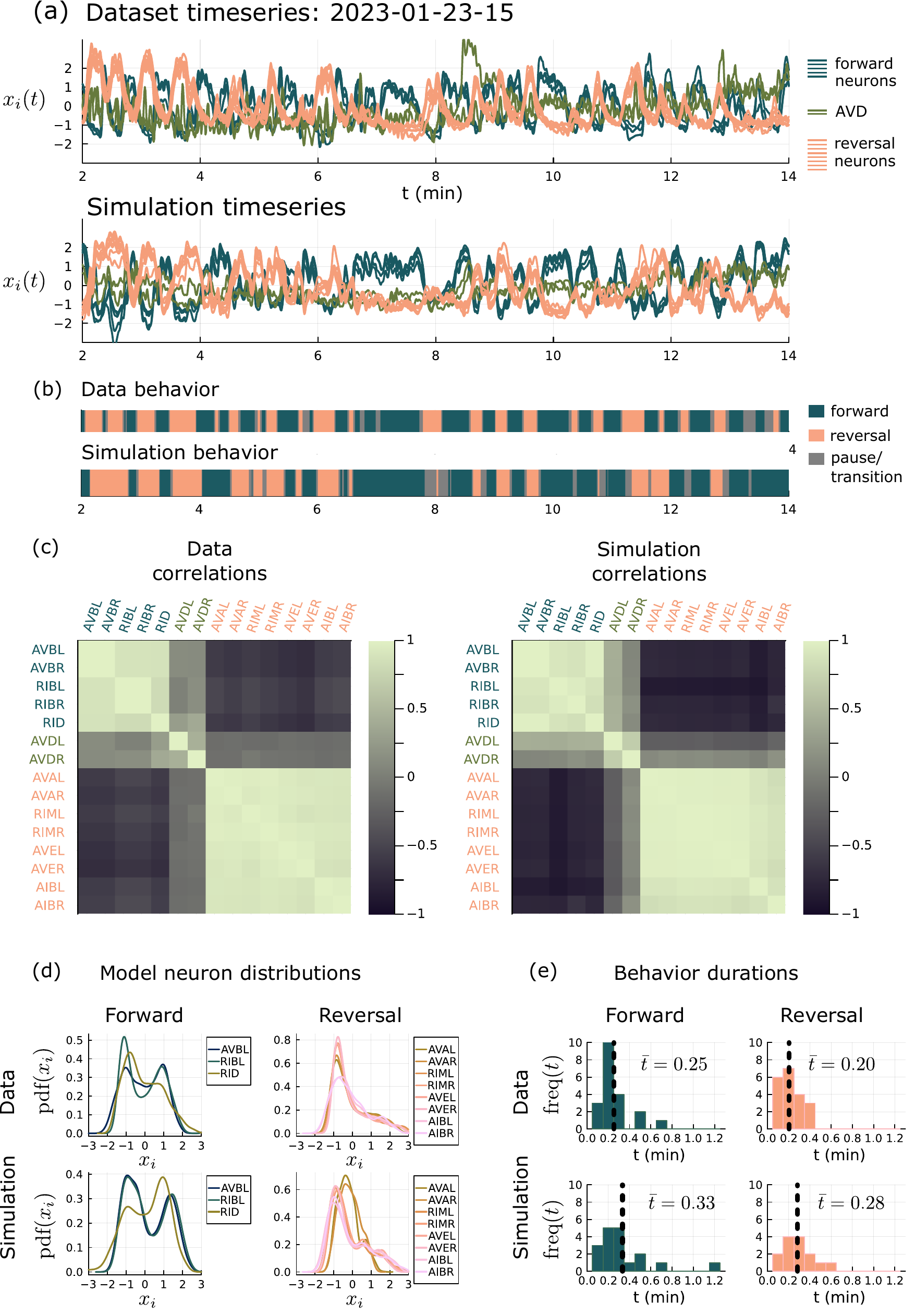}
   \caption{Comparison of data and model. (a) Core neuron activity in the dataset versus the simulation. (b) The dominant cluster of neurons --- forward or reversal --- is a proxy for behavior (see text for detail). (c) Pairwise correlations between neurons. (d) Distribution of neural activity.  (e) Dwell times in the forward and reversal states. Histograms are generated from the forward/reversal sequences in (b) using bin size $0.1$ and excluding dwell times less than $0.05$ min. }
   \label{fig:higher_order_stats}
\end{figure}

In this section we compare model outputs to data, demonstrating that though
much simplified, our model can reproduce many key features of \textit{C. elegan} locomotion.

\paragraph{\large Switching dynamics}
A salient characteristic of \textit{C. elegans} locomotion is a constant switching between forward and reversal movements at characteristic yet variable time intervals.  An appropriate first test of our premotor network model is thus to evaluate if the model can reproduce these switching dynamics.
In Figure~\ref{fig:higher_order_stats} we compare premotor neuron activity from dataset 2023-01-23-15.json \cite{atanas_brain-wide_2023} to simulated activity of the corresponding neurons.
Of interest are the activity levels of neurons in the core group; $x_i(t)$ from this dataset will be compared to simulation results from our model when driven by signal neurons whose outputs are taken from the same dataset.
Figure~\ref{fig:higher_order_stats}(a) shows $x_i(t)$ for individual neurons over a 12-minute duration: the top panel is data from \cite{atanas_brain-wide_2023}, the bottom panel is the simulation results.
In both cases, neurons in the forward and reversal clusters, and the AVD neurons, are shown in three different colors, and the partial synchronization within each group is visible.
Qualitatively, the forward and reversal clusters in the simulation switch between high and low states similar to the switches seen in the data.

Assuming that a high state of the forward cluster represents forward locomotion and the analogous statement for the reversal cluster, Figure~\ref{fig:higher_order_stats}(b) shows the forward/reversal direction of movement for this \textit{C. elegans} as a function of time.
We let $F(t)$ be $x_i(t)$ averaged among the forward neurons and let $R(t)$ be $x_i(t)$ averaged among the reversal neurons.
We assume that when $F(t)-R(t)> 0.5$ (dark green) the \textit{C elegans} is
engaging in forward movement as, on average, the forward neurons are more active than the reversal neurons.  When $F(t)-R(t)<-0.5$  (tan) we assume it is in reversal as the reversal neurons are more active. 
For $|F(t)-R(t)| < 0.5$ (shown in grey), we consider the movement to be ambiguous 
or paused. The two strips showing data and simulations are 
remarkably similar given the many simplifying assumptions made in the design of
our premotor network model.  More examples of comparison data and model comparisons are shown in Figure~\ref{fig:simulation_data_match}.

\paragraph{\large Higher order statistics}
Figure~\ref{fig:higher_order_stats}(c) shows the correlations between pairs of neurons. 
Data and simulation compare well: 
Neurons in the forward cluster are highly correlated 
and anticorrelated to the reversal neurons. An analogous statement holds for neurons in the reversal cluster. The AVD neurons are not strongly correlated with either cluster in both the data and simulation.

Figure~\ref{fig:higher_order_stats}(d) shows the probability distribution functions of the activity levels for the neurons in question.  Here as well, model statistics resemble data: e.g. the forward neurons are bimodal in their pdfs and the reversal neurons have a single strong peak below zero and a large positive tail.

Figure~\ref{fig:higher_order_stats}(e) shows the dwell time distributions for the forward and reversal states.  The histograms are qualitatively similar, with mean dwell times between 0.2 and 0.33 minutes in both data and simulations. Simulated dwell times are slightly longer, but note that the time constant $\tau$ in our model is optimized over all data sets.

 We stress that the parameters in our network model are deduced from many datasets and are not specific to the dataset 2023-01-23-15. Indeed, our model simulates forward/reversal switching that resembles the actual switching in many datasets (Fig.~\ref{fig:simulation_data_match}). 

 \subsection*{Analysis} 

\paragraph{\large Gap junctions versus synaptic connections}

\begin{figure}[th!]
   \centering
   \includegraphics[width = 1.0\linewidth]{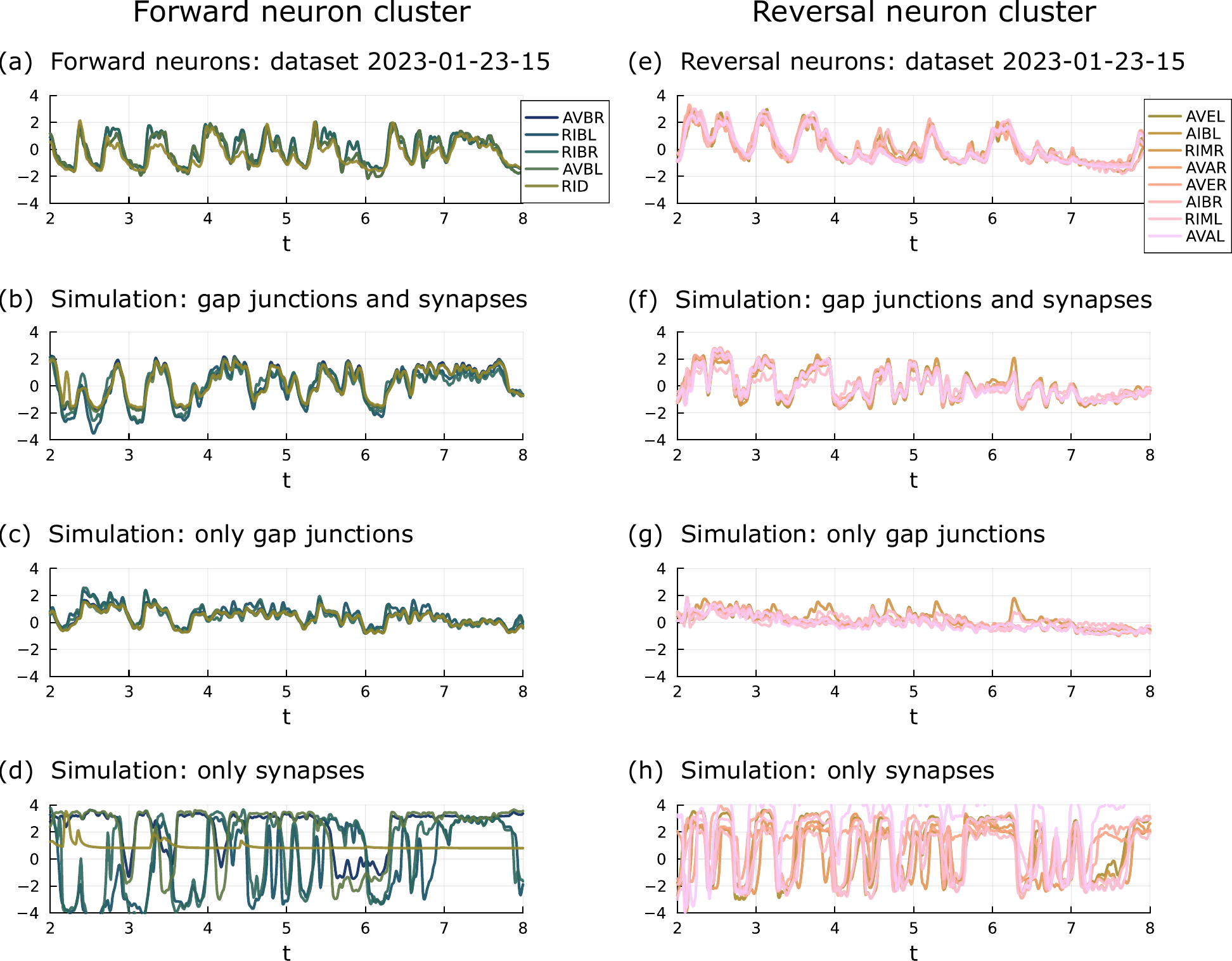}
   \caption{Forward and reversal cluster dynamics using connected neurons as signals.  (a) Forward neuron activity from dataset 2023-01-23-15. (b) Simulated forward neuron activity using both gap junctions and synapses. (c) Simulated forward neuron activity using only gap junctions. (d) Simulated forward neuron activity using only synapses.  (e)  Reversal neuron activity from dataset 2023-01-23-15.  (f) Simulated reversal neuron activity using both gap junctions and synapses. (g) Simulated reversal neuron activity using only gap junctions. (h) Simulated reversal neuron activity using only synapses.}
   \label{fig:avb_ava_sim}
\end{figure}

Next, we use our premotor network model to examine the roles played by gap junctions versus synapses in the stochastic switching between forward and reversal seen in Figure~\ref{fig:higher_order_stats}.
First, we observe two salient characteristics in the dynamics of the forward
neurons. Figure~\ref{fig:avb_ava_sim}(a) shows the time series from a data set. 
Figure~\ref{fig:avb_ava_sim}(b) shows the corresponding time series, simulated using our
premotor network model when driven by signal neurons from the same dataset.
Note that in the data as in the model, the forward neurons are synchronized, and
 together they make
irregular but characteristic switches between high and low states.

Next, we study the effect of removing the synapses in the model, leaving only gap junctions.
Fig.~\ref{fig:avb_ava_sim}(c) shows that the forward neurons stay correlated but 
fail to produce clear, strong switches between the high and low states. 
When gap junctions are removed leaving only synaptic interaction, 
the forward neurons exhibit dramatic switches between states but do not stay 
synchronized (Fig.~\ref{fig:avb_ava_sim}(d)).

Similar results are observed for the reversal cluster (Fig.~\ref{fig:avb_ava_sim}(e)-(h)).

These results show that in our premotor network, both gap junctions and synaptic connections are needed to produce the switching behavior observed in data.
They play different roles: gap junctions synchronize, while the stochastic switching appears to be mediated by synaptic dynamics. We hypothesize that similar mechanisms may be at work in the real \textit{C. elegans} brain.


\paragraph{ \large Behavior over longer time durations: roaming {\it vs} dwelling}
 While semi-regular switching between forward and reversal is characteristic
of \textit{C. elegans} locomotion, the patterns and frequencies of these switches are known 
to convey information on the behavior of the worm over longer time durations. 
Roaming and dwelling are two such complementary behavioral states. 
Roaming allows \textit{C. elegans} to explore a large space to find food while dwelling 
keeps \textit{C. elegans} in a more confined region so that it can,  e.g.,
 exploit a discovered 
food source \cite{flavell_behavioral_2020}. Roaming is characterized by long stretches 
of forward runs interrupted by infrequent reversals while dwelling is characterized by 
shorter forward runs and frequent, short, reversals \cite{flavell_behavioral_2020}.

\begin{figure}[ht!]
   \centering
   \includegraphics[width = 1.0\linewidth]{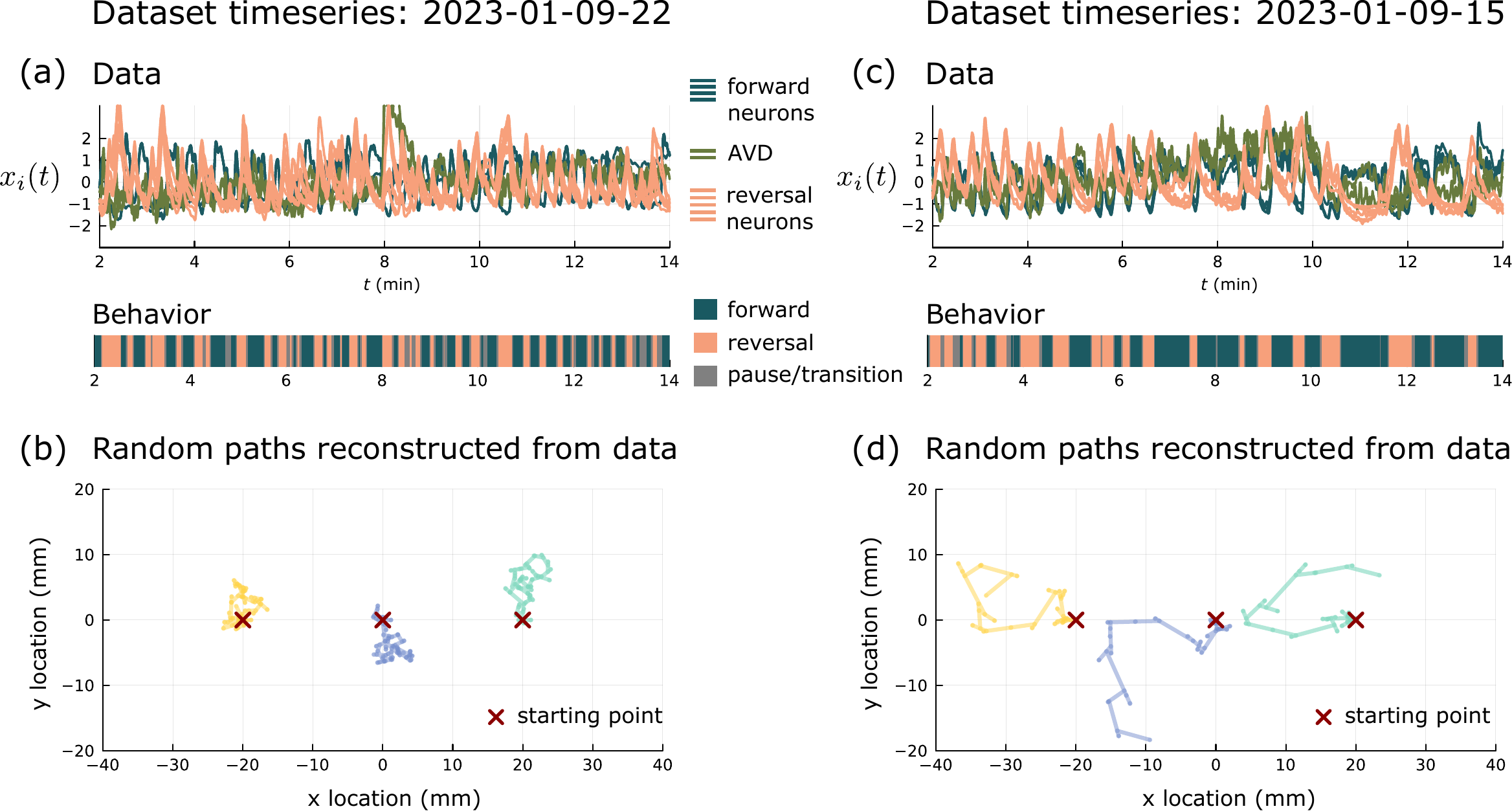}
   \caption{(a) Core neuron activity in dataset 2023-01-09-22 \cite{atanas_brain-wide_2023} and resulting locomotory state time series. (b) Three locomotion paths simulated using the locomotion time series in (a) and the simulation procedure described in the \hyperref[methods:loco]{Methods}.  (c-d) Analogous results for dataset 2023-01-09-15.}
   \label{fig:roaming_vs_dwelling}
\end{figure}

Here we propose that it may be possible to infer these \textit{C. elegans} behaviors from the activity time series 
of just a handful of neurons, such as the forward and reversal clusters used in our model. 
Figure~\ref{fig:roaming_vs_dwelling} contrasts dwelling and roaming behavior from 
two Atanas et al.~(2023) datasets: 2023-01-09-22 on the left column and 2023-01-09-15 on the right \cite{atanas_brain-wide_2023}.
Figure~\ref{fig:roaming_vs_dwelling} (a) and (c) show the time series of the forward
and reversal neurons and summary forward/reversal transitions (as in Fig 3(b)). 
We then construct locomotory paths that simulate \textit{C. elegans} locomotion 
on an agar plate: Following \cite{zhao_reversal_2003}, a simplified pattern of forward-run, 
reversal, turn, and resume-forward is imposed. The durations of forward and reversal are from 
the summary transition panels shown in Fig.~\ref{fig:roaming_vs_dwelling}(a) and (c). Forward and reversal speeds are fixed and turn angles are randomly selected from a distribution.
 Different draws of turn angles 
result in different realizations of the locomotory path, three of which are shown for each
of the two datasets (Figure~\ref{fig:roaming_vs_dwelling} (b) and (d)). More details on
path reconstruction are given in the \hyperref[methods:loco]{Methods}.

In the dataset on the left (Fig.~\ref{fig:roaming_vs_dwelling}), the switching occurs at higher frequencies, resulting in shorter forward
runs and reversals. As a result, the \textit{C. elegans} stays confined to a smaller region 
over the 12-minute time interval (Fig.~\ref{fig:roaming_vs_dwelling}(b)); this behavior is characteristic of dwelling. The dataset 
on the right has longer forward runs and longer reversals, resulting in all three simulated 
paths in Figure~\ref{fig:roaming_vs_dwelling}(d) exploring somewhat larger regions than in 
Figure~\ref{fig:roaming_vs_dwelling}(b). These paths show characteristics in the
direction of roaming.
Locomotory paths constructed from simulated transitions (as in 
Figure~\ref{fig:higher_order_stats}(b)) give similar results.


\begin{figure}[th!]
   \centering
   \includegraphics[width = 1.0\linewidth]{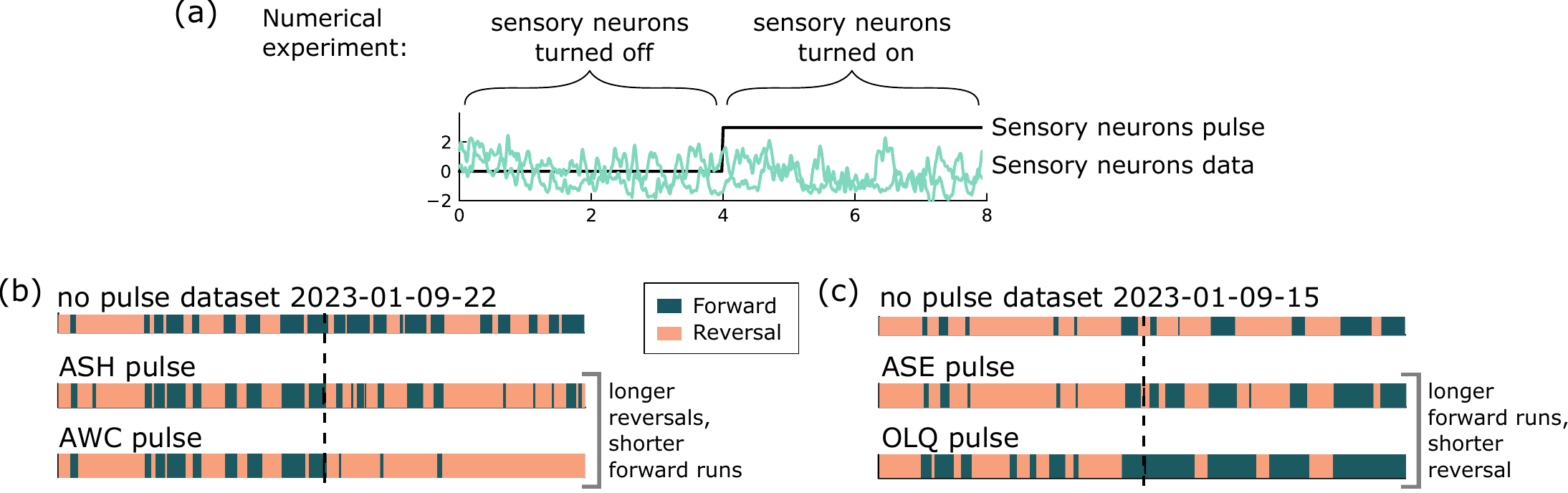}
\caption{(a) In each numerical experiment, a Heaviside pulse time series replaces the dataset time series for a sensory neuron class.
(b) The behavior measured from the simulated core neuron activity exhibits longer reversals when receiving pulses from the ASH and AWC sensory neuron classes.  (c) The behavior measured from the simulated core neuron activity exhibits longer forward runs when receiving pulses from the ASE and OLQ sensory neuron classes.}
\label{fig:neurons_pulse}
\end{figure}

\paragraph{\large Manipulation of sensory inputs}
 One of the goals of building a model is to avail ourselves of
a tool to explore how different sensory inputs affect behavior. 
This is a high bar: for a model to give reliable results, it must be biologically
very realistic. Here we evaluate our model's responses to individual sensory
inputs. Specifically, we select a few signal neurons in the model,
modify their activity to be a Heaviside pulse (see  Fig.~\ref{fig:neurons_pulse}(a)), and
compare the effect on model dynamics to these neurons' known functions as determined in the experimental literature.

Figure~\ref{fig:neurons_pulse}(b) shows the results of manipulating signal neurons
ASH and AWC. Experimental findings show that the ASH sensory neurons initiate an avoidance response; they sense touch to the nose as well as noxious chemicals and in response trigger the termination of forward motion and the initiation of reversals \cite{bargmann_chemosensation_2006, kaplan_dual_1993, sambongi_sensing_1999}. 
AWC\textsuperscript{ON} sensory neurons help \textit{C. elegans} navigate chemical gradients by triggering turns in response to negative derivatives \cite{itskovits_concerted_2018}. Consistent with the experimental literature, our regression
results show that ASH and AWC have excitatory synapses with the reversal cluster (Fig.~\ref{fig:syn_weights_avgs}(b)), 
suggesting that they should be reversal-promoting. Unsurprisingly, when we activate either of these
two neurons in our simulation, reversals become more frequent and prolonged 
compared to the unperturbed simulation  (Fig.~\ref{fig:neurons_pulse}(b)).

More ambiguous outcomes are reported in Figure~\ref{fig:neurons_pulse}(c).
Experiments have shown that ASE neurons facilitate chemotaxis to water-soluble attractants \cite{bargmann_chemosensory_1991, bargmann_chemosensation_2006};
ASEL lengthens bouts of forward motion while ASER promotes turns \cite{suzuki_functional_2008}. 
In our model, ASE sensory neurons are found to be connected to the reversal cluster with synaptic weights
close to zero  (Fig.~\ref{fig:syn_weights_avgs}(b)).
Activating them slightly lengthens the forward runs
(Figure~\ref{fig:neurons_pulse}(c)),  a result not inconsistent with the data.
Our last example is the OLQ neurons, which form gap junctions with forward neurons only
and do not have synaptic connections with the core group. It follows logically that their activation 
would be forward-promoting, a fact confirmed in model simulations:
forward runs are very prolonged (Figure~\ref{fig:neurons_pulse}(c)). In reality, however,
 the function of OLQ is complex: it is implicated 
in head withdrawal and nose touch avoidance and it supports foraging \cite{riddle_mechanosensory_1997};
and when IL1 and OLQ neurons are ablated, \textit{C. elegans} forage abnormally slowly \cite{riddle_mechanosensory_1997}. We hypothesize that these different behaviors supported by OLQ 
may be carried out through different pathways not captured by our model.

In summary, these tests show that our model has the capability to make some
predictions correctly, but it needs to be upgraded to capture more complex functions.

\section*{Discussion}\label{sec:discussion}

\subsubsection*{\large Relation to existing literature}

The functions of individual \textit{C. elegans} neurons and their association with behavior have been extensively studied \cite{kato_global_2015, flavell_behavioral_2020, pirri_neuroethology_2012, lanza_recurrent_2021, flavell_serotonin_2013, zheng_neuronal_1999}.
Experimentalists have identified interneurons that are primarily responsible for driving and facilitating the primary locomotion behaviors of forward crawling, reversals, and turns \cite{zhen_c_2015, wakabayashi_neurons_2004, piggott_neural_2011}.
Ablation studies show that key premotor neurons are essential for producing forward and reversal movements \cite{chalfie_neural_1985, wicks_integration_1995}.
Whole-brain imaging shows that many neurons beyond the command premotor neurons are correlated with primary locomotion behaviors;  moreover, network activity as a whole encodes behaviors better than the activity of individual neurons \cite{kato_global_2015, atanas_brain-wide_2023}.
Many additional neurons and circuits have been linked to specific higher-order behaviors such as roaming and dwelling, local and global search, quiescence, chemotaxis, and the escape response \cite{flavell_behavioral_2020, suzuki_functional_2008, itskovits_concerted_2018, gray_circuit_2005, pirri_neuroethology_2012}. 
Still, the question remains: 

\medskip
 {\it How do these neurons work together to determine behavior, sustain different behaviors, and switch between them?}

\medskip \noindent
This study aims to shed light on these questions.

Many phenomenological models have been presented for the seemingly random
switching between forward and reversal neurons which underlies \textit{C. elegans} locomotion.
Roberts et al.~(2016) reproduced stochastic switching with a Markov model and proposed mutual inhibition and inherent stochasticity as the mechanisms for switching \cite{roberts_stochastic_2016}.
Linderman et al.~(2019) proposed a model with random switches between discrete states each described by linear dynamics \cite{linderman_hierarchical_2019}.  Chaotic heteroclinic networks can reproduce switching statistics, highlighting that deterministic neural dynamics can generate seemingly random state switches \cite{morrison_chaotic_2022}.
Kato et al.~(2015) observed that motor commands are represented globally and found that the neural representation of motor sequences evolved on a low-dimensional manifold obtained from PCA \cite{kato_global_2015}.
Fieseler et al.~(2020) built a low-dimensional model of this neural activity actuated by control signals that induce switches \cite{fieseler_unsupervised_2020}. %
Morrison et al.~(2021) built a similar model using
nonlinear dynamics \cite{morrison_nonlinear_2021-1}.
These models connect switching phenomena to mathematical frameworks, but it is hard to draw biological inferences from them as they are too far removed from neuroanatomy.


 Other authors have studied biophysical network models of premotor neurons. Rakowski et al.~(2013) and (2017) simulated the dynamics of a pre-motor and motor circuit. The stationary distributions of the motor neurons were then used to infer synaptic polarities, i.e., whether a synaptic
connection is excitatory or inhibitory \cite{rakowski_synaptic_2013, rakowski_optimal_2017}; 
there is little discussion of network dynamics. Lanza et al.~(2021) simulated the dynamics of the \textit{C. elegans} neuronal network and evaluated the activity of the command neurons \cite{lanza_recurrent_2021, mcculloch_logical_1943}. Their model predicted that neural activity converges to limit cycles; i.e., all neurons eventually acquire the same periodicity.
Previous biophysical network models assigned polarities to synapses through other means and explored the behavior of premotor neurons when activated by one or two specific sensory neurons \cite{kunert_spatiotemporal_2017, kim_neural_2019, gleeson_c302_2018}; 
%
%
Ref.~\cite{kunert_spatiotemporal_2017} found also that their network dynamics
converged to a limit cycle.

Our biophysical network model presents a different dynamical picture,
one that does not fit with classical paradigms such as limit cycles (which
are too simplistic given that forward-reversal transitions occur at irregular times). As our core system
is driven by over a hundred signal neurons with diverse functionalities, our simulated
premotor dynamics are complex and varied, with seemingly random
switching statistics matching what is observed in imaging data.
We have benefited from many extensively labeled whole-brain imaging datasets which became
available only recently and which we used to fit model parameters \cite{atanas_brain-wide_2023}.
These datasets have enabled us to build a model that is directly connected to
anatomy and has more realistic model outputs.

Because our model is semi-realistic, it has the potential to shed light on how connection type, structure, and signal input contribute to \textit{C. elegans} locomotion.
We find, for example, that gap junctions and synaptic connections support stochastic switching in different ways. 
We observe how different switching statistics can result in qualitative differences in simulated locomotory paths.  Our model also shows how elevated input rates from even single classes of neurons can significantly alter switching statistics and therefore \textit{C. elegans} behavior.

\subsubsection*{\large Challenges in building a mechanistic network that is data-driven, biophysical, and analyzable}

 One of the challenges in building such a model is the extraction of a suitable subnetwork. \textit{C. elegans} neurons are highly recurrently connected; they do not process information in a feed-forward manner. Many premotor neurons are known to be ``hubs" in the network with extensive connections \cite{uzel_set_2022}. These network features obfuscate which subset of neurons are responsible for determining locomotion -- yet analyzability
demands clarity and simplicity. 
An innovation in this paper is to select a core group of neurons of manageable size and model their dynamics as they receive input from a larger group of neurons (called signal neurons) the outputs of which we glean directly from data.
We propose that approximating networked systems as a core system driven by external input can be a paradigm for modeling complex network dynamics in mathematical biology.

While it is far more realistic than studying a closed circuit driven by
one or two input sources, the approach we have introduced is not without limitations: To keep the size of the core group manageable, we have had to omit a significant amount of feedback from the core neurons to the signal neurons. Because we view the premotor neurons as primarily driving the motor neurons, we have neglected feedback from motor neurons.

Our model is also constrained by data availability. We do not include neurons for which we lack whole-brain imaging data.  Specifically, we exclude the PVC and DVA neurons despite their strong recurrent connections with the core set.  Many additional neurons are rarely or never labeled in whole-brain imaging data which means that their contribution to the dynamics is not represented \cite{atanas_brain-wide_2023}.
We expect that the accuracy of the model would improve if these neurons could be included.

\begin{figure}[th!]
    \centering
    \includegraphics[width = 1.0\linewidth]{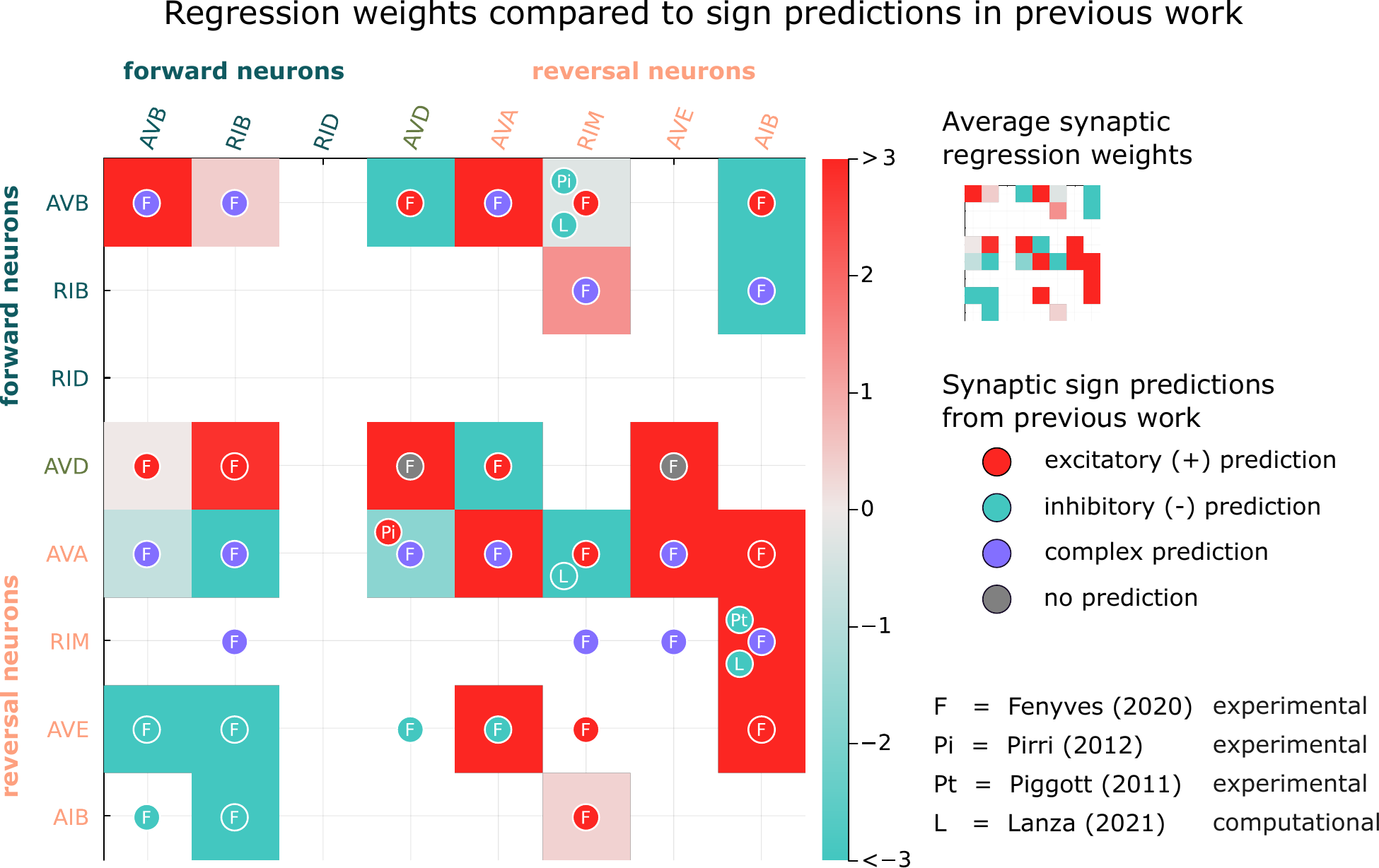}
    \caption{Polarities obtained from regression compared to polarities determined from previous work \cite{fenyves_synaptic_2020, pirri_neuroethology_2012, piggott_neural_2011, lanza_recurrent_2021}. Synaptic weights are grouped and averaged over each neuron class (e.g. AVBL and AVBR and combined to form AVB). 
 Ref.~\cite{fenyves_synaptic_2020} determines polarities for synapses that are not in the Ref.~\cite{white_structure_1986} dataset which is used to establish the location of synapses for the model in this study.}
    \label{fig:syn_weights_scatters}
\end{figure}

A major challenge in modeling biological systems is integrating disparate and inconsistent experimental evidence (and/or inferences from other computational studies).
An example of the uncertainty we have encountered is in
the signs of synaptic weights in the \textit{C. elegans} connectome.
%
For example, Fenyves et al.~(2020) contains an extensive set of experimentally determined polarity predictions using gene expression data \cite{fenyves_synaptic_2020}. 
Though many of the polarities are labeled as complex due to the existence of both excitatory and inhibitory neurotransmitter-receptor pairs between the pre and postsynaptic neurons, overall they predict more excitatory than inhibitory connections (Fig.~\ref{fig:syn_weights_scatters}).
In contrast, Rakowski et al.~(2017) predicts, based on a computational model, that most interneurons in the circuit controlling locomotion (AVA, AVB, AVD, AVE, DVA, PVC) are inhibitory \cite{rakowski_synaptic_2013, rakowski_optimal_2017}.  
 Not all experimental studies agree on sign predictions either.
For example, the synaptic connection from RIM to AVB is excitatory according to Fenyves et al.~(2020) but inhibitory according to Pirri et al.~(2012) (Fig.~\ref{fig:syn_weights_scatters}) \cite{fenyves_synaptic_2020, pirri_neuroethology_2012}.

Because different studies predict different synaptic polarities, it is unclear how to use previous studies to constrain model parameters; our predictions were obtained independently through regression based on whole-brain data. Nor is it possible -- given these discrepancies -- to match all existing data.
Figure~\ref{fig:syn_weights_scatters} shows synaptic signs and weights between core neurons in our model 
 (grid box color) compared to the polarities predicted by several previous studies.
As can be seen, our synaptic predictions 
match some previous predictions and deviate from others.  
For example, we predict inhibitory synapses from forward neurons AVB and RIB to reversal neurons AVE and AIB, and mostly excitatory synapses from the reversal neurons to other reversal neurons,
matching fairly well the findings of Ref.~\cite{fenyves_synaptic_2020}. On the other hand, RIM to AVA is inhibitory in our model and Ref.~\cite{lanza_recurrent_2021} but excitatory in Ref.~\cite{fenyves_synaptic_2020}.

 We remark that despite the discrepancies in polarity predictions, our model reproduces stochastic switching behavior, suggesting that such behavior is very robust.

\subsubsection*{\large Ongoing and future work}

While our model captures transitions between the two primary locomotory states, forward and reversal, there are several other locomotory states whose neural correlates have been identified, such as dorsal and ventral turns, omega turns, and quiescence \cite{kato_global_2015, linderman_hierarchical_2019, nichols_global_2017}. 
The methods developed in this paper can be used to study these states.
Our model can also be used to examine the impact of the environment on premotor activity through sensory neurons. Signals from the environment such as oxygen levels and chemical gradients
are critical for chemotaxis and escape responses.

A biologically semi-realistic model such as ours can shed light on
{\it dynamic mechanisms}, including mechanisms for sustaining and switching
between different locomotion states, and for higher-order behavioral states such as dwelling and roaming.  
 It can also be used to analyze the effects of ablations and identify the function of signal neurons with respect to locomotion.

Going forward, the driven dynamical systems paradigm proposed in this paper has many
potential applications. Closer to this work is the modeling and analysis of other circuits in 
the \textit{C. elegans}  nervous system, and the integration of
circuits for multiple functions into more comprehensive models.
Further afield, the techniques developed here can be exported to other
model organisms with available connectomic or
large-scale imaging data.
%


\section*{Methods}\label{sec:methods}

\normalsize

\bigskip
In the \hyperref[sec:results]{Results} section, we pose a general model for \textit{C. elegans} neural dynamics constructed using the connectome and experimentally observed intrinsic dynamics, and then fit model parameters using whole-brain imaging data.  Here we fill in details omitted in the main text, mainly on (a) the modeling of intrinsic dynamics
of individual neurons, and (b) how some of the parameters are fitted. We also provide more details  on (c) network simulations 
such as those  in Figure~\ref{fig:higher_order_stats}, and (d) our construction of locomotion paths in Figure~\ref{fig:roaming_vs_dwelling}.

\subsection*{Model dynamics and parameters}

As stated in the \hyperref[sec:results]{Results} section, each of the core neurons, $\vec{\b{x}} = [x_1, x_2,..., x_n]^T \in \mathds{R}^n$, has the following approximate dynamics for its calcium imaging brightness $x_i$ (neuron GCamp (z-scored)):
\begin{align}\label{eq:general_model}
\tau \frac{dx_i}{dt} &=  f_i(x_i) + \beta \sum_{j=1}^{n+m}\b{W}_{ij}(x_j - x_i) +\sum_{j=1}^{n+m}\b{A}_{ij} \sigma(x_j)
\end{align}
where $f(x_i)$ is neuron $x_i$'s intrinsic dynamics, $\beta \sum_{j=1}^{n+m}\b{W}_{ij}(x_j - x_i)$ represents the influence from neurons that form gap junctions with neuron $x_i$, and $\sum_{j=1}^{n+m}\b{A}_{ij} \sigma(x_j)$  represents the influence from neurons that are presynaptic to neuron $x_i$. $\sigma() = \text{ReLU}()$ is the ReLU activation function and $\tau$ is a global timescale variable.
The dynamics of the signal neurons are not modeled, their dynamics are extracted from the whole-brain imaging time series.


\subsubsection*{Intrinsic dynamics}\label{method:intrin_d}

\begin{figure}[ht!]
    \centering
    \includegraphics[width=0.9\linewidth]{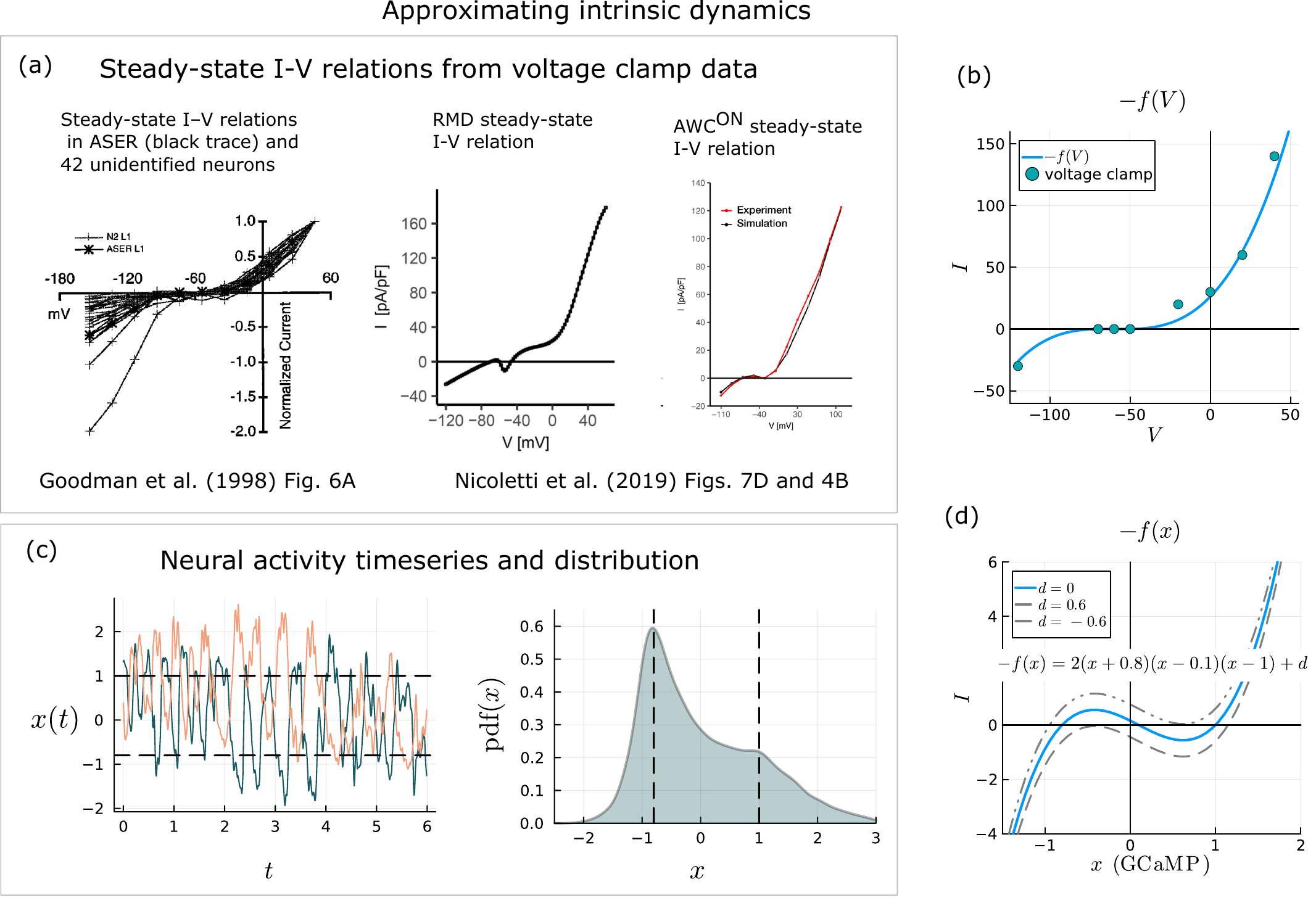}
    \caption[Caption for LOF]{(a) Voltage clamp experiments from Ref.~\cite{goodman_active_1998}. This figure has been reproduced with permission \footnotemark.
    Voltage clamp experiments from Ref.~\cite{nicoletti_biophysical_2019}.  (b) Intrinsic dynamics approximated from voltage clamp data in terms of voltage, $V$. (c) Calcium imaging time series and distribution over premotor neurons. The pdf of the time series of the premotor neurons is used to set fixed points for the intrinsic dynamics, $x_{fp} = -0.8, 0.1, 1$.  (d) Negative of approximate intrinsic dynamics, $-f(x)$, in terms of the calcium imaging variable, $x$.}
    \label{fig:voltage_clamp}
\end{figure}

Voltage and current clamp experiments show that \textit{C. elegans} neurons have a graded, nonlinear response to input current \cite{goodman_active_1998, nicoletti_biophysical_2019}. Due to voltage-dependent ion channels, they are highly sensitive to input around their resting membrane potential and become less sensitive to input far from their resting membrane potential (Fig.~\ref{fig:voltage_clamp}(a)).
Suppose that the dynamics for a neuron's voltage is $\frac{dV}{dt} = f(V)$. An approximation for $f(V)$ can be attained through voltage clamp experiments. The dynamics for a neuron's voltage when an input current is applied to the neuron is $\frac{dV}{dt} = f(V) + I$. Voltage clamps stabilize a neuron's voltage at a given level $V$, essentially setting the derivative equal to zero, $\frac{dV}{dt} = 0$, by applying a current $I$ that will satisfy this objective. 
The amount of current $I$ necessary to stabilize the neuron at a range of voltage levels $V$ provides an approximation for $-f(V)$.
We approximate $-f(V)$ for the inter/motor neuron RMD with the cubic function $-f(V) = \frac{1}{8000}(V+70)(V+60)(V+50)$ (Fig.~\ref{fig:voltage_clamp}(b)).
This cubic function approximately matches the RMD steady-state I-V relation derived in Ref.~\cite{nicoletti_biophysical_2019} (Fig.~\ref{fig:voltage_clamp}(a)).
The approximate function for intrinsic voltage dynamics, $f(V) = -\frac{1}{8000}(V+70)(V+60)(V+50)$, indicates that there are two stable fixed points in the voltage dynamics at approximately $V = -70$ and $V = -50$ and an unstable fixed point at $V = -60$. This bistability is corroborated by experiments that show that RMD neurons have a plateau potential at approximately $-35$ mV in addition to its resting membrane potential at $-70$ mV \cite{lockery_quest_2009}.

While voltage clamp experiments can provide approximate forms for the intrinsic dynamics in terms of voltage, we need formulas for the intrinsic dynamics in terms of calcium imaging brightness.
While there is no formula for the relationship between voltage and calcium imaging for \textit{C. elegans} neurons, we can estimate the average resting membrane potential and sensitivity range in $x$, the neuron GCamp variable, by taking the distribution of $x$ over the premotor neurons and finding peaks in the distribution (Fig.~\ref{fig:voltage_clamp}(c)). We find the most common value across neurons is $x = -0.8$, and so set this as the average resting membrane potential (Fig.~\ref{fig:voltage_clamp}(c)). There is another bump in the distribution around $x=1$, and so we set this to be the plateau potential stable state in the intrinsic dynamics with an unstable fix point between the two stable fixed points at $x = 0.1$ (Fig.~\ref{fig:voltage_clamp}(d)). The resulting cubic $f(x) = -2(x+0.8)(x-0.1)(x-1)$ is a qualitative estimate for the intrinsic dynamics of interneurons in general.
The locations of the fixed points and their stabilities vary from neuron to neuron. Voltage clamps show that while the intrinsic dynamics of neurons have the same qualitative shape, the fixed point locations differ \cite{goodman_active_1998, nicoletti_biophysical_2019} (Fig.~\ref{fig:voltage_clamp}(a)). We include the bias term $\b{d}_i$ in the intrinsic dynamics to account for this variability (Fig.~\ref{fig:voltage_clamp}(d)).

We set the intrinsic dynamics of each neuron in terms of $x_i$, without input from other neurons to be
\begin{align}
\frac{dx_i}{dt} &= f_i(x_i) = -2(x_i+0.8)(x_i-0.1)(x_i-1)  + \b{d}_i
\end{align}
where $\b{d}_i$ varies from neuron to neuron. When $\b{d}_i = 0$, the intrinsic dynamics have two stable states at $x_i = -0.8$ and $x_i=1$ and one unstable state at $x_i = 0.1$. 
Allowing a different bias term $\b{d}_i$ for each neuron allows the locations and stability of the fixed points to vary across neurons while keeping the same general form.  Depending on the value of $\b{d}_i$, one of the stable fixed points can become semi-stable or eliminated (Fig.~\ref{fig:voltage_clamp}(d)).

The cubic form we use to approximate the intrinsic dynamics captures several key properties.
Firstly, it allows for voltage bistability in interneurons. Plateau potentials allow neurons to have the same computational properties as Schmitt triggers \cite{lockery_quest_2009, mellem_action_2008}. Neurons that lack two stable fixed points due to the bias term still have richer computational properties than if they possessed linear intrinsic dynamics. 
Second, the cubic form acts to provide bounds on the voltage range of each neuron. Neurons do not have an unlimited capacity for hyperpolarization or depolarization; negative feedback pressures keep the voltage within a bounded range and our model captures this constraint with the negative cubic term $-x^3$.
Saturation effects must be included to ensure the model produces realistic neural activity.
Thirdly, the cubic form reflects the neurons' increased sensitivity to input near its resting membrane potential and plateau potential. The neurons' sensitivity to input varies in a voltage-dependent manner which is consistent with experimental data \cite{goodman_active_1998}.

The cubic form we use for the intrinsic dynamics of the \textit{C. elegans} neurons has similarities to the voltage dynamics in models for spiking neurons.
Both the FitzHugh–Nagumo and Hindmarsh–Rose models of neuron dynamics approximate the voltage dynamics with a cubic function \cite{fitzhugh_impulses_1961, nagumo_active_1962, hindmarsh_model_1997}.  One of the key differences between these models and our model is that models of spiking neurons incorporate recovery variables that capture spiking and the instability of the high-voltage state. \textit{C. elegans} neurons do not spike and can be stable at a high-voltage state so we approximate each neuron's intrinsic dynamics without using recovery variables \cite{lockery_quest_2009, mellem_action_2008}.

\subsubsection*{Gap junctions and synaptic weights}\label{method:gap_j}

The second term in Eq.~\ref{eq:general_model}, $ \beta \sum_{j=1}^{n+m}\b{W}_{ij}(x_j - x_i)$, captures the influence of neurons connected to neuron $x_i$ via gap junctions.
$\b{W}$ is a symmetric matrix that holds the relative strengths of gap junctions between pairs of neurons; these weights are taken from the connectome \cite{white_structure_1986}. The weights in $\b{W}$ are applied to the difference between $x_j$ and $x_i$, meaning that the activity of neurons connected via gap junctions are pressured to equalize.
The entire term is scaled by $\beta$ which determines the relative contribution of gap junctions to the overall dynamics; we fit $\beta$ with a parameter sweep over many fits with whole-brain imaging data.

\footnotetext{Reprinted from \textit{Neuron}, 20(4), Miriam Goodman, David Hall, Leon Avery, and Shawn R Lockery, \textit{Active currents regulate sensitivity
and dynamic range in C. elegans neurons}, 763–772, Copyright (1998), with permission from Elsevier.}


\begin{figure}[th!]
    \centering
    \includegraphics[width = 1.0\linewidth]{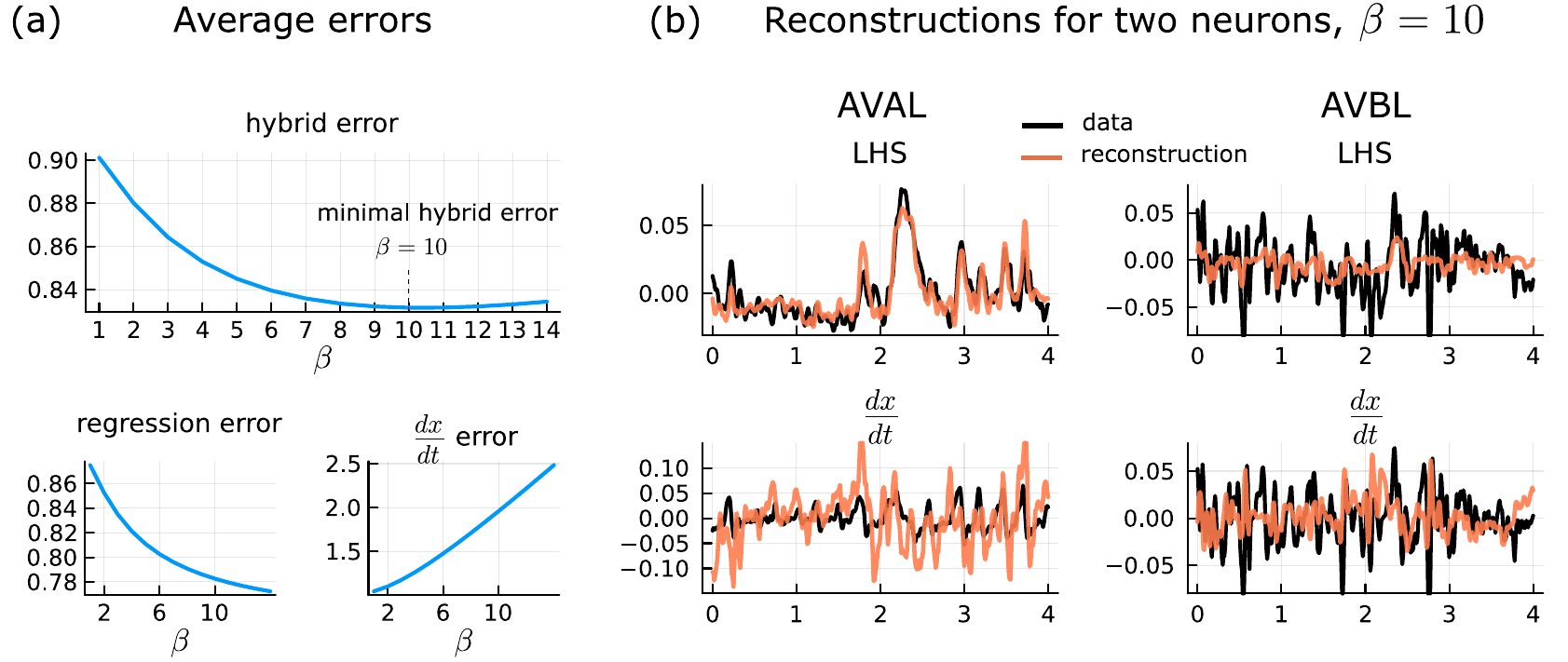}
    \caption{(a) Average hybrid error as a function of $\beta$. The average error is a combination of the regression error and the derivative reconstruction error.  
    Derivative and reconstruction errors are the L2 norm of the difference between the LHS and RHS of Eqs.~\ref{eq:general_model} and \ref{eq:dyn_sys_wsetparams} respectively.
 (b)  Example regression LHS and derivative reconstruction for AVAL and AVBL for $\beta = 10$. }
    \label{fig:error_param_sweep}
\end{figure}

The third term in Eq.~\ref{eq:general_model}, $\sum_{j=1}^{n+m}\b{A}_{ij}\sigma(x_j)$ captures the input from neurons presynaptic to neuron $x_i$.
$\b{A}$ is a nonsymmetric matrix that contains the signed, weighted, and directed synaptic inputs between pairs of neurons. 
The edge graph for synaptic connections in \textit{C. elegans} is derived from \cite{white_structure_1986}.

\subsubsection*{Parameter fit}\label{method:param_fit}

We determine $\beta$, $\b{A}$, and $\b{d}$ simultaneously by performing multiple linear regressions to approximate $\b{A}$ and $\b{d}$ for different values of $\beta$ across 22 datasets selected from Ref.~\cite{atanas_brain-wide_2023}. Our criteria for the selection of these 22 datasets is that the dataset must contain at least one of the following forward core neurons --- AVBL, AVBR, RIBL, or RIBR --- because they are sparsely labeled in the datasets and one neuron in each class of the reversal neurons and AVD.  To reduce the error from missing neurons we perform data replacement on missing time series when possible by substituting missing time series with the time series of highly correlated neurons. Certain pairs of left/right neurons are consistently highly correlated and therefore one neuron's time series can be used as a proxy for the other if one neuron in the pair is missing. We label neuron pairs as highly correlated if they are on average at least 70\% correlated in the datasets where they both appear; 30 neurons fit this criteria. For these highly correlated neurons, when one of the neurons is missing we use the time series of the other neuron as a proxy in the regression.

We set $\tau = 1$ and fit the remaining parameters.
From Eq.~\ref{eq:general_model} we subtract the default intrinsic dynamics and gap junction terms from the derivatives, leaving the synaptic weights and bias terms on the right-hand side, 
\begin{align}
\label{eq:dyn_sys_wsetparams}
    \frac{dx_i}{dt} +2(x_i+0.8)(x_i-0.1)(x_i-1) - \beta \sum_{j=1}^n\b{W}_{ij}(x_j - x_i) &= \sum_{j=1}^n\b{A}_{ij} \sigma(x_j) + \b{d}_i.
\end{align}
Eq.~\ref{eq:dyn_sys_wsetparams} provides us with a reformulation of Eq.~\ref{eq:general_model} where all parameters on the left-hand side (LHS) are determined with the exception of $\beta$. We fix different values for $\beta$ and then perform linear regressions for the 22 datasets to approximate $\b{A}$ and $\b{d}$ using snapshots from the calcium imaging time series of all neurons $\b{x}(t)$ acquired from Ref.~\cite{atanas_brain-wide_2023}.

We perform a parameter sweep for $\beta$ and select the $\beta$ value that minimizes a combination of the regression error and the reconstructed derivative error across all datasets and across core neurons (Fig.~\ref{fig:error_param_sweep}(a)). 
The regression error, $\text{error}_{regress}$, is the L2 norm of the difference between the timseries of the left-hand side and right-hand side of Eq.~\ref{eq:dyn_sys_wsetparams}. The reconstructed derivative error, $\text{error}_{deriv}$, is the norm of the difference between the time series of the derivative and the reconstruction of the derivative, Eq.~\ref{eq:general_model}.
The hybrid error is a combination of the regression error and the derivative error, $\text{error}_{hybrid} = \text{error}_{regress} + 0.05 * \text{error}_{deriv}$.
The average regression error, derivative error, and hybrid error are shown for varying $\beta$ values in Figure~\ref{fig:error_param_sweep}(a). As an example of error for individual neurons, the top row of Figure~\ref{fig:error_param_sweep}(b) shows the time series data of the left-hand side of Eq.~\ref{eq:dyn_sys_wsetparams} with its reconstruction for neurons AVAL and AVBL.  The bottom row shows the time series data of the derivatives of AVAL and AVBL along with the derivative reconstructions, Eq.~\ref{eq:general_model}.

We observe the hybrid error across datasets and core neurons is minimized at $\beta = 10$ (Fig.~\ref{fig:error_param_sweep}).
We set $\beta = 10$ and then perform linear regressions for $\b{A}$ and $\b{d}$ estimates across the 22 datasets.  We use the average of the weights and biases across all datasets as the synaptic weights and biases in the general model.

\paragraph{Whole-brain imaging datasets}

We use calcium imaging time series data from Ref.~\cite{atanas_brain-wide_2023}.
Each dataset consists of calcium imaging time series data for $64-106$ labeled neurons \cite{atanas_brain-wide_2023}. Access to this data is provided through the Worm Wide Web database: \href{https://wormwideweb.org/dataset.html}{https://wormwideweb.org/dataset.html}.

\subsubsection*{Premotor network simulation}\label{methods:sim}

We simulate the dynamics of core neurons by integrating Eq.~\ref{eq:general_model}, using the time series of signal neurons as input. 
We perform numerical integration in Julia using the Runge-Kutta-Fehlberg (RKF45) method.
We fit the timescale parameter by finding $\tau$ such that the simulated neural dynamics have the same reaction times as those observed in the neural recordings across the datasets shown in Figure~\ref{fig:simulation_data_match}. 
The closest reaction time results from $\tau = 0.2$. 
Because each dataset contains traces for only a subset of neurons, each simulation is performed with input from a subset of the theoretical input. To help compensate for the missing input, we amplify the existing synaptic input by a magnification factor. We simulate the dynamics using increasingly large magnification factors and find a magnification factor of 1.4x brings the core neurons to the correct activity level.

\begin{figure}[th!]
    \centering
    \includegraphics[width = 0.7\linewidth]{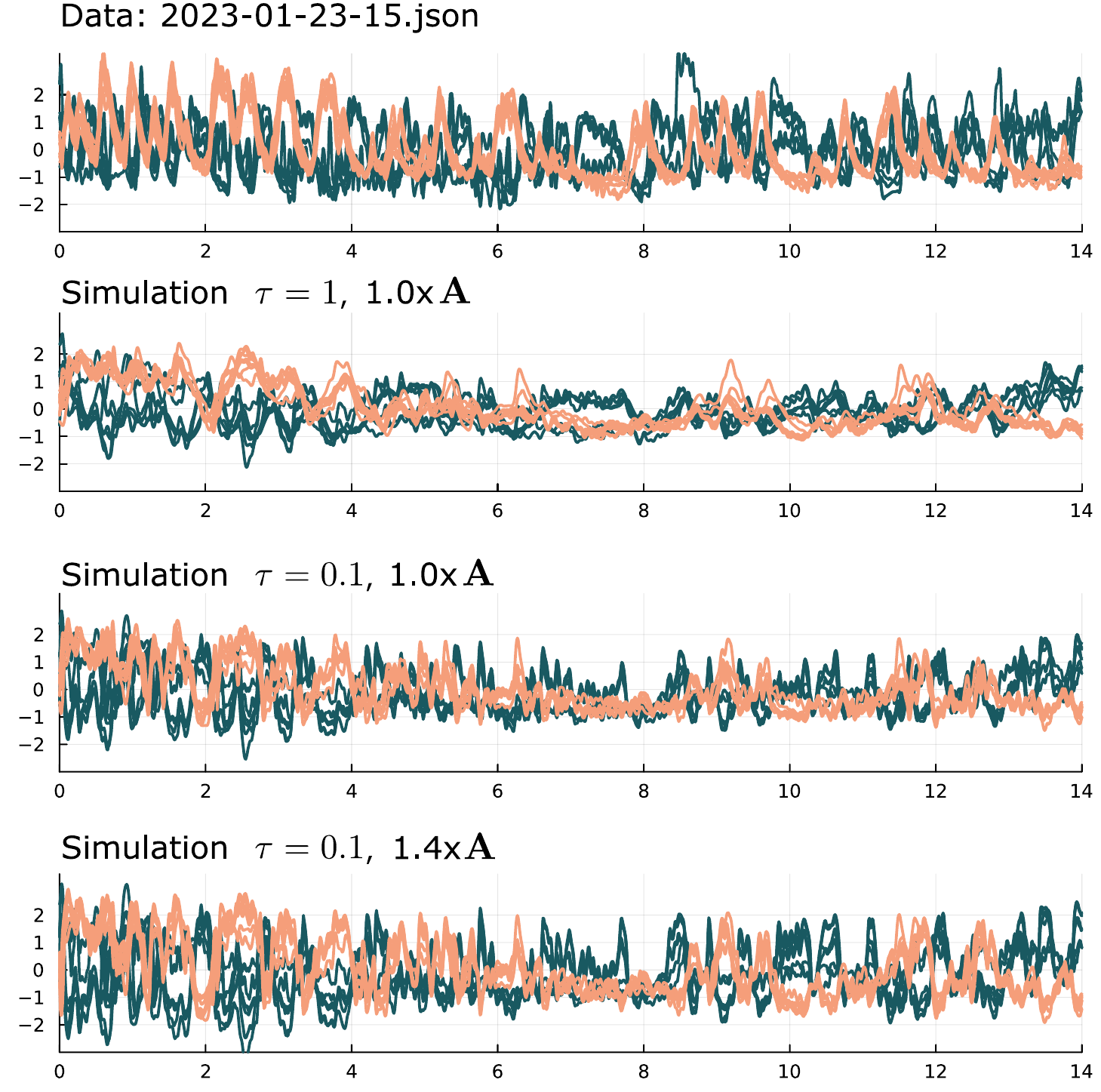}
    \caption{Dataset compared to simulations with different timescales and magnification factors.  Decreasing the timescale parameter $\tau$ makes the system respond quicker to stimuli. Because the system is receiving partial signals we also magnify $\b{A}$ by 1.4x. Increasing the amount of stimulus to the premotor neurons increases their average activity levels.}
    \label{fig:timescale_increase}
\end{figure}

\begin{figure}[th!]
    \centering
    \includegraphics[width = 1.0\linewidth]{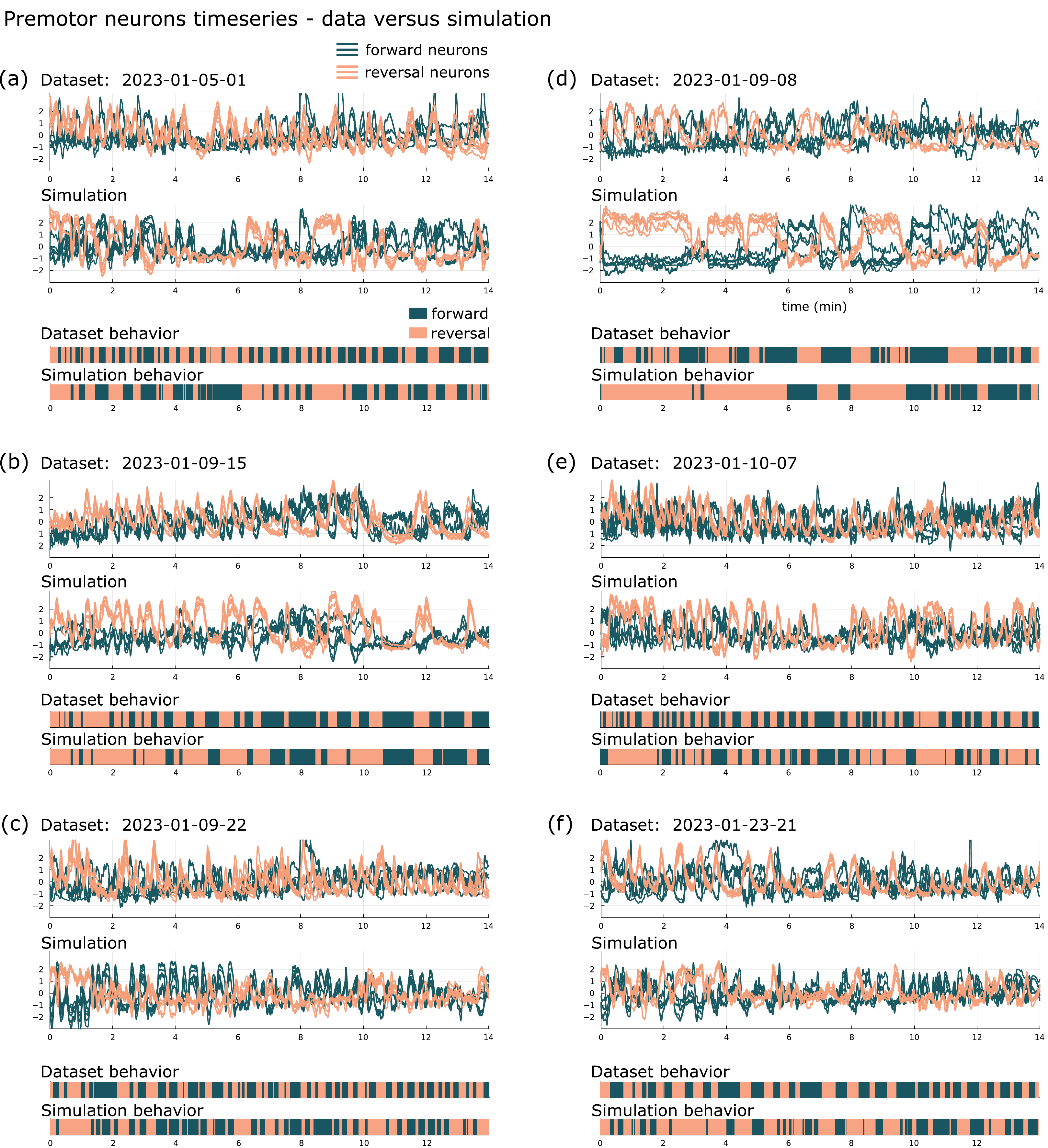}
    \caption{time series of core neurons in whole-brain imaging data compared to simulation. Behavior is measured as the dominant cluster (forward or reversal). $\tau = 0.2$, magnification on synaptic input is 1.4.}
    \label{fig:simulation_data_match}
\end{figure}

An example of how $\tau$ and the magnification factor affect the simulation is shown in Figure~\ref{fig:timescale_increase}.
Neural recordings from premotor neurons in dataset 2023-01-23-15 are compared to simulations performed with varying timescale parameters $\tau$ and magnification factors for the synaptic input (Fig.~\ref{fig:timescale_increase}).
Decreasing the timescale parameter $\tau$ makes the system react quicker to input.
Increasing the magnification factor increases the amount of input applied to the neurons which results in higher average activity levels in the neurons.  This phenomena is similarly observed in the other datasets.
By decreasing the timescale parameter as well as magnifying the input from $\textbf{A}$ we get simulated behavior that more closely matches the premotor neural recordings. 

Using a single set of parameters, the general model reproduces the core neural activity observed in many datasets (Fig.~\ref{fig:simulation_data_match}).  While the parameter values remain the same, the input provided to the core neurons varies from dataset to dataset, resulting in different network activity.

\subsection*{Locomotion path simulation}\label{methods:loco}

We simulate example locomotion paths for the forward/reversal behavior sequences, obtained from the data, using a simplified version of the procedure outlined in Zhao et al.~(2003)~\cite{zhao_reversal_2003}.  Three sample paths for each dataset are initialized at the red marks.  The locomotion cycle consists of a forward run, a reversal, a turn following the reversal, and then the resumption of forward motion.
The forward speed is set to $0.15$ \si{mm/sec} and the reversal speed to $0.075$ \si{mm/sec}.  The probability of a regular turn is $65\%$ while the probability of an omega turn is $35\%$.  Regular turn angles are selected from a uniform distribution ranging from $-90 \degree$ to $90 \degree$.  Omega turns are deeper turns; they are selected from a uniform distribution ranging from $90 \degree$ to $270 \degree$.  Because turn angles are random, the three simulated paths for each dataset differ despite having the same forward/reversal sequences and durations.

\subsection*{Data Availability Statement}

There are no primary data in the paper.  The model weights and source code to reproduce the simulations and statistical analysis is available on GitHub at \url{https://github.com/mmtree/Celegans_premotor}.

\section*{Acknowledgments}\label{sec:ack}
This work was supported by the National Science Foundation (MM, award no.~2103239 and LSY, grants DMS-1901009 and DMS-2350184).  The majority of this work was done while MM was an NSF Mathematical Sciences Postdoctoral Research Fellow at the Courant Institute of Mathematical Sciences at New York University.

\end{document}